# Physical Insights of Low Thermal Expansion Coefficient Electrode Stress Effect on Hafnia-Based Switching Speed

Y.-T. Tsai[1], C.-R. Liu[1], Y.-T. Chen[1], S.-M. Wang[1], Z.-K. Chen[1], C.-S. Pai[1], Z.-R. Haung[1], F.-S. Chang[2], Z.-X. Li[2], K.-Y. Hsiang[2], M.-H. Lee[2], Y.-T. Tang[1]*
[1] Dept of Electrical Engineering, National Central University
No. 300, Zhong-da Rd., Zhongli District, Taoyuan City 320317, Taiwan (R.O.C.)
[2] Institute of Electro-Optical Engineering, National Taiwan Normal University, Taipei 11677, Taiwan
Phone: +886-972-815-521 E-mail: yttang@ee.ncu.edu.tw

***Abstract***

In this report, we investigate the effect of low coefficient of thermal expansion (CTE) metals on the operating speed of hafnium-based oxide capacitance. We found that the cooling process of low CTE metals during rapid thermal annealing (RTA) generates in-plane tensile stresses in the film, This facilitates an increase in the volume fraction of the o-phase and significantly improves the domain switching speed. However, no significant benefit was observed at electric fields less than 1 MV/cm. This is because at low voltage operation, the defective resistance (dead layer) within the interface prevents electron migration and the increased RC delay. Minimizing interface defects will be an important key to extending endurance and retention.

## ***Introduction***

Hafnium oxide is a polarizable material like the early piezoelectric materials. When an electric field is applied, in addition to the generation of dipole, the unstable oxygen element also shifts. After the electric field is dropped, the insulator maintains the polarization of the electric dipole. Due to the asymmetrical phase, hafnium-based oxides possess the binary and can be fabricated into bits. This property makes ferroelectric materials practical for many applications, including ferroelectric memory, ferroelectric capacitors, and neuromorphic computing. Take FeFET for example, it operates fast (<50ns) and has an endurance of $10^9$, which is close to the nanosecond of DRAM with endurance $10^{12}$. In recent years, TSMC, Intel, Micron and other leading IC companies have focused on developing various storage class memories (SCM). In 2022 IEDM, Intel proposed AFE ferroelectric memory, whose high K value (60) and long endurance (~$10^{11}$) have been shown to replace DRAM as the last cache near the CPU [1]. In addition, 3D-IC vertical integration boosts the storage density and expands the applications. To achieve this goal, it becomes increasingly important to find technologies compatible with BEOL channels, such as IGZO for 3D-FeNAND. In general, the leakage of metal oxides becomes severe above 500°C. To possess ferroelectricity and multibit at < 450°C is much more challenging in fabrication process. In this work, we investigate how to meet this goal with low CTE gate level metals.

To understand the switching speed of HZO, we used molybdenum (Mo) as the capping layer to compare TiN/Mo/HZO/TiN with TiN/HZO/TiN MFM structures. The switching of the HZO domain was driven under a series of pulsed voltages of different lengths and the energy required for the switching of HZO was extracted. Finally, it was demonstrated that the electron migration rate of HZO under the Mo capping electrode was faster than that under the TiN capping electrode.

## ***Analysis & Result***

For comparing the effect of Mo capping layer and TiN capping layer on the reversal rate of ferroelectric polarization, in the first step, we cleaned the Si substrate via RCA and then deposited TiN as the bottom electrode by sputtering. Then, we ALD ferroelectric HZO:9nm (1:1 ratio of $HfO_2$ to $ZrO_2$) on the TiN electrode with process temperature of 250°C. Then, we sputtered different top electrodes and pattern 80 µm × 80 µm FeCAP structures by using i-Line exposure and dry etching to create, followed by annealing at 450 °C for 30 seconds. The process flow is shown in Figure 1.

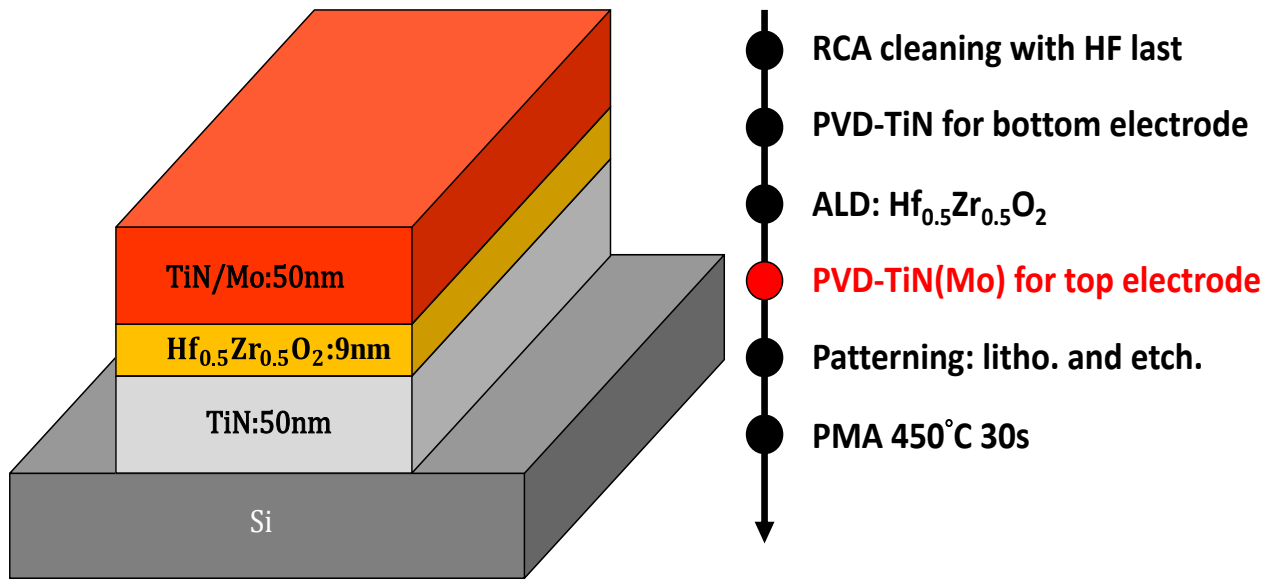

**Fig.1** Process flow and FeCAP structure

Regarding the measurements, a radiometric ferroelectric measurement instrument was used to evaluate the PV curve of FeCAP, as shown in Figure 2. Under operating frequency of 1kHz, the FeCAP switching steepness (dP/dV) of the Mo capping layer is 3 times higher than that of the traditional TiN. In addition, the remannat polarization 2Pr value is observed to be 46 µC/cm2, which is 1.3 times higher than that of conventional TiN electrodes.

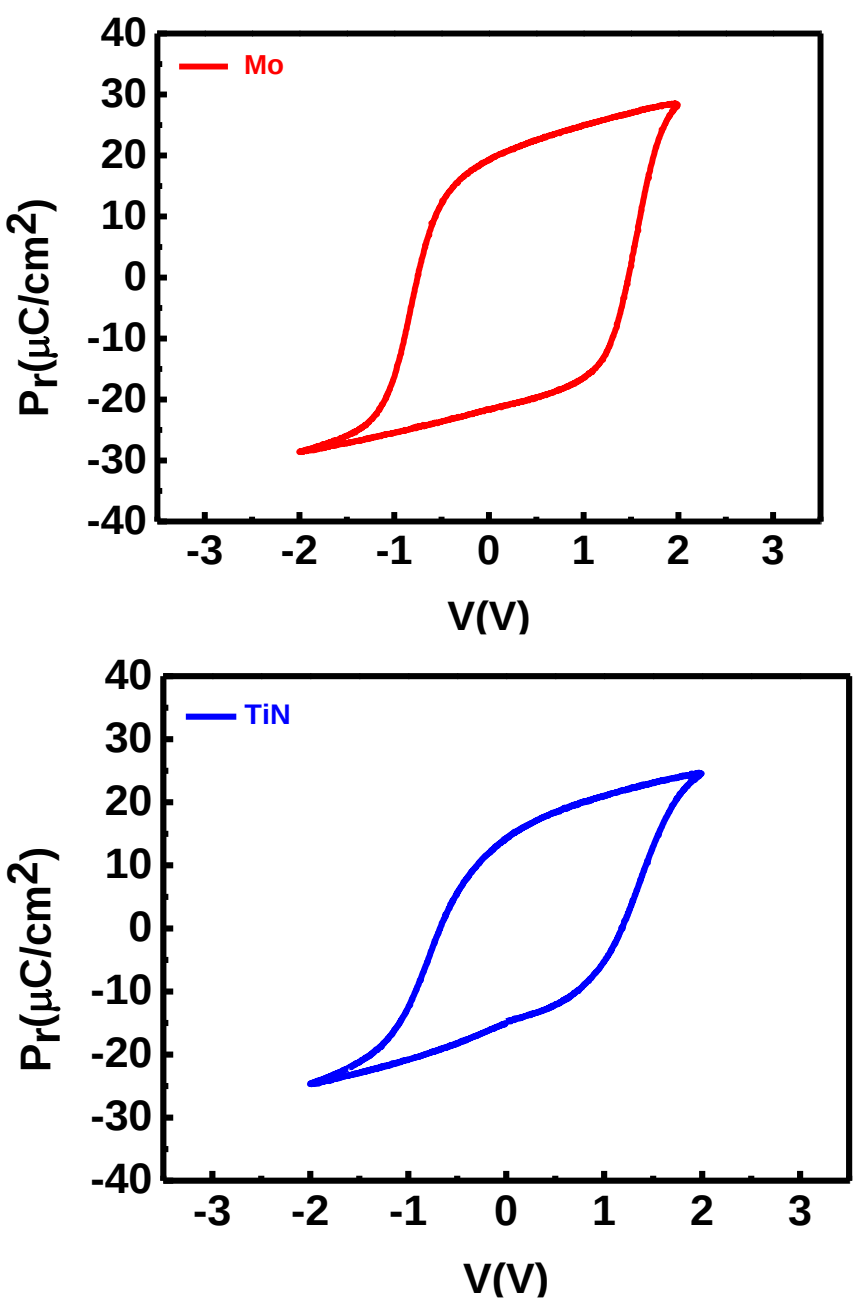

**Fig.2** Comparison of PV curves with different top electrodes

The read/write speed of FeCAP is determined by the polarization switching limit. To measure the read/write speed limit, we used Keysight 1530A to generate PUND and measured the polarization switching behavior of 80μm×80μm FeCAP at different pulse lengths ($t_p$). The input signal is shown in Fig. 3(a), and the polarization of Mo-cap and TiN-cap are shown in Figs. 3(b) and 3(c), respectively. We observe that the $\Delta P/\Delta t_p$ of Mo-capping is steeper than that of TiN-capping, while the voltage of Mo-capping generating reverse polarization domains is smaller than that of TiN-capping at the same degree of polarization. To explain this phenomenon, we show the relationship between the polarization switching speed and the electric field in Fig. 4.

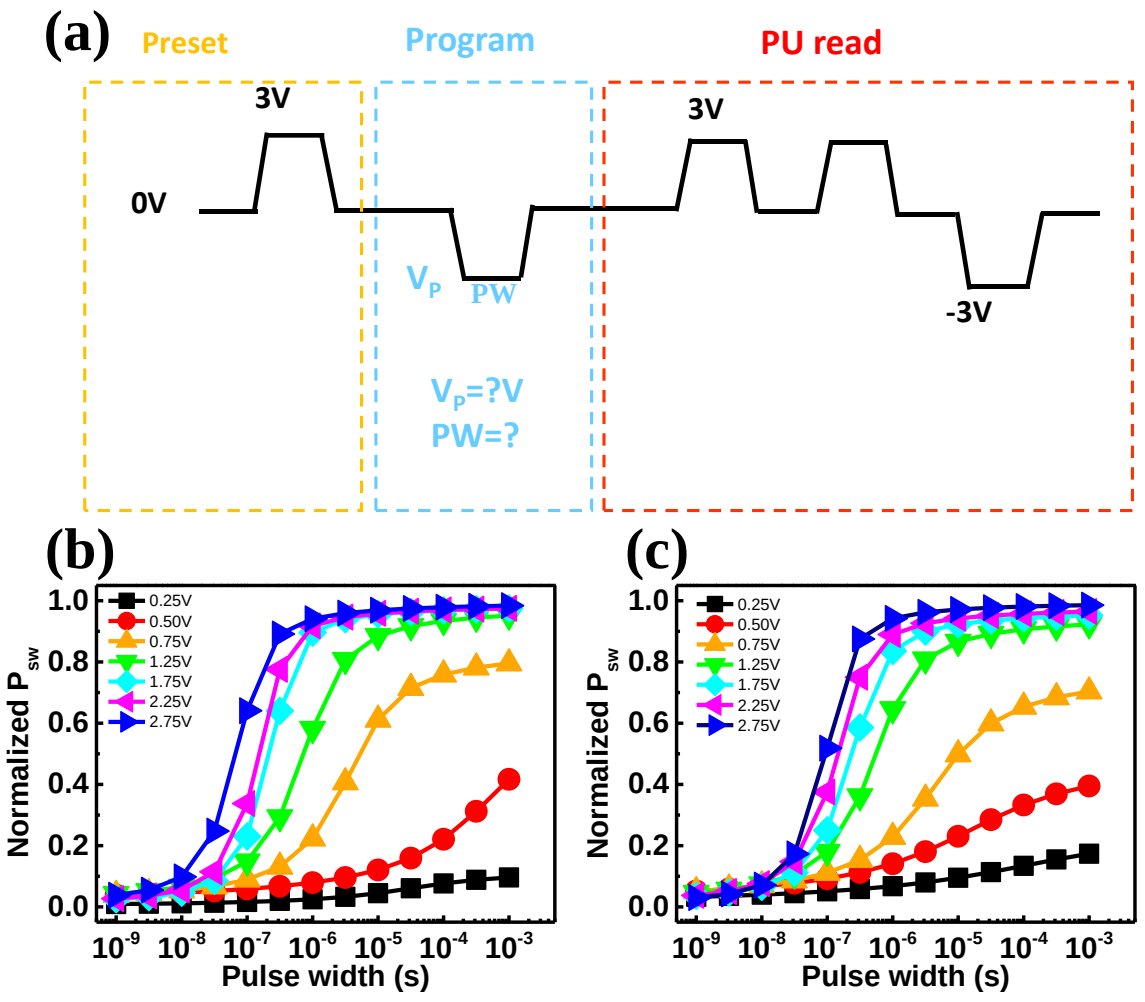

**Fig.3 (a)** pulse sequence for the PUND measurement **(b)** polarization degree vs. pulse for Mo-capping **(c)** polarization degree vs. pulse for TiN-capping.

Figure 4 summarizes the relationship between the switching speed and the operating electric field and finds that when the electric field is >1 MV/cm, the domain dominates the ferroelectricity. The electric field drives the domain switching, and the o-phase volume content within the domain determines the polarization. According to ref. [2], low CTE electrodes form in-plane tensile stress on the insulator during RTA cooling process, facilitating t-phase to o-phase. The CTE of Mo (TiN) is about 6(9) ($10^{-6}\cdot K^{-1}$), which is much lower than that of HZO ($10\cdot 10^{-6}\cdot K^{-1}$), and the Mo electrode annealing provides a total volume change of 4% (as shown in Fig. 4(a)). This enables Mo-capping to generate a rich o-phase (ferroelectricity). Furthermore, Fig. 4(b) shows a plot of operating speed versus electric field, and it can be found that the read/write speed exhibits linear growth for electric fields from 1 to 2.25 V and nonlinear growth above 2.5 V, we ascribes this phenomena to the grain size. On the contrary, at <1 MV/cm, the polarization is no longer determined by the domain, but by the electronic dipole. For low voltage, TiON shows stronger polarity than metallized $MoO_x$, so the start-up speed is faster.

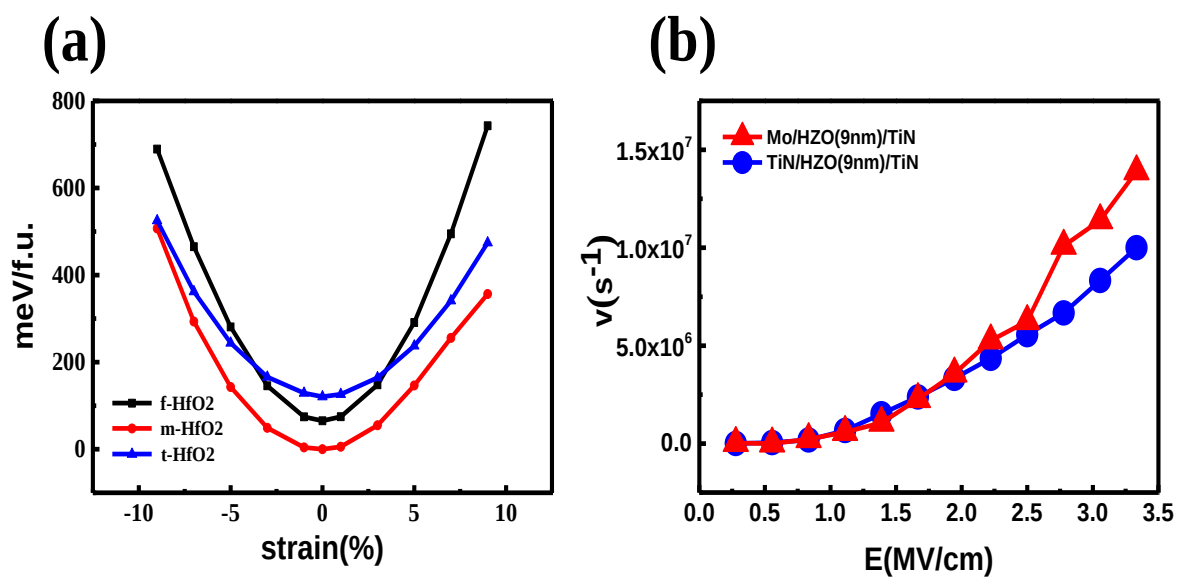

**Fig.4 (a)** Calculated Gibbs energy of the f-/t-phase of $HfO_2$ relative to the m-phase at different stress, where the f-phase and t-phase undergo phase transition at tensile strain 4% **(b)** Relationship between Polarization Switching Speed and Applied Electric Field.

## Conclusion

In this work, we use pulse measurements to study the polarization switching speed under different kinds of gates. We found that metals with low coefficient of thermal expansion (CTE) can generate in-plane tensile stress in the film during cooling process of RTA, which helps to increase the volume fraction of o-phase and remarkably improve the domain polarization switching speed. However, this benefit is not expected because electron dipole dominates at low voltage operation.

## Acknowledgement

This work was financially supported by the NSTC 111-2218-E-A49-016-MBK, NSTC 111-2622-8-A49-018-SB, NSTC 109-2221-E-008-093-MY3.